\documentclass{acmtog}
\usepackage{amssymb}
\usepackage{amsmath}
\usepackage{hyperref}

\acmVolume{VV}
\acmNumber{N}
\acmYear{YYYY}
\acmMonth{Month}
\acmArticleNum{XXX}  
\acmdoi{XX.XXXX/XXXXXXX.YYYYYYY}

\begin{document}

\markboth{A. Mukhopadhyay et al.}{Detection and Characterization of Intrinsic Symmetry}

\title{Detection and Characterization of Intrinsic Symmetry} 

\author{ANIRBAN MUKHOPADHYAY {\upshape and} SUCHENDRA M. BHANDARKAR
\affil{Department of Computer Science, The University of Georgia, Athens, GA 30602-7404, USA}
\and 
FATIH PORIKLI 
\affil{Mitsubishi Electric Research Laboratories, Inc., 201 Broadway, Cambridge, MA 02139, USA }}


\terms{Symmetry, Intrinsic Geometry}



\maketitle

\begin{bottomstuff} 

\end{bottomstuff}

\begin{abstract} 
  
A comprehensive framework for detection and characterization of overlapping intrinsic symmetry over 3D shapes is proposed. To identify prominent symmetric regions which overlap in space and vary in form, the proposed framework is decoupled into a {\it Correspondence Space Voting} procedure followed by a {\it Transformation Space Mapping} procedure. In the correspondence space voting procedure, significant symmetries are first detected by identifying surface point pairs on the input shape that exhibit {\it local} similarity in terms of their intrinsic geometry while simultaneously maintaining an intrinsic distance structure at a {\it global} level. Since different point pairs can share a common point, the detected symmetric shape regions can potentially overlap. To this end, a {\it global intrinsic distance-based voting} technique is employed to ensure the inclusion of only those point pairs that exhibit significant symmetry. In the transformation space mapping procedure, the {\it Functional Map} framework is employed to generate the final map of symmetries between point pairs. The transformation space mapping procedure ensures the retrieval of the underlying dense correspondence map throughout the 3D shape that follows a particular symmetry. Additionally, the formulation of a novel cost matrix enables the inner product to succesfully indicate the complexity of the underlying symmetry transformation. The proposed transformation space mapping procedure is shown to result in the formulation of a semi-metric symmetry space where each point in the space represents a specific symmetry transformation and the distance between points represents the complexity between the corresponding transformations.  Experimental results show that the proposed framework can successfully process complex 3D shapes that possess rich symmetries.

\end{abstract}

\section{Introduction}

Symmetry is ubiquitous in nature and in artificial man-made  objects. The detection and characterization of shape symmetry has attracted much attention in recent times, especially within the computer graphics community~\cite{Mitra13}. Although most of the existing literature has focused on the detection of {\it extrinsic} symmetries; a popular approach being transformation space voting~\cite{Mitra06}; there has been steadily growing interest in detection and characterization of {\it intrinsic} symmetries. Most recent efforts in intrinsic symmetry detection have focused on detection of global symmetries~\cite{Ovsjanikov08};~\cite{Lipman10}. It is generally recognized that detection of overlapping intrinsic symmetry is a more challenging problem due to the larger search spaces involved in the detection of symmetric regions (in comparison to global symmetry analysis) and the determination of symmetry revealing transforms (in comparison to extrinsic symmetry detection). Also, overlapping intrinsic symmetry detection and characterization is more important because it is a more generalized problem in nature than extrinsic symmetry detection problem which can be considered as a special case of overlapping intrinsic symmetry detection. 


\begin{figure*}
\begin{center}
   \includegraphics[width=150mm]{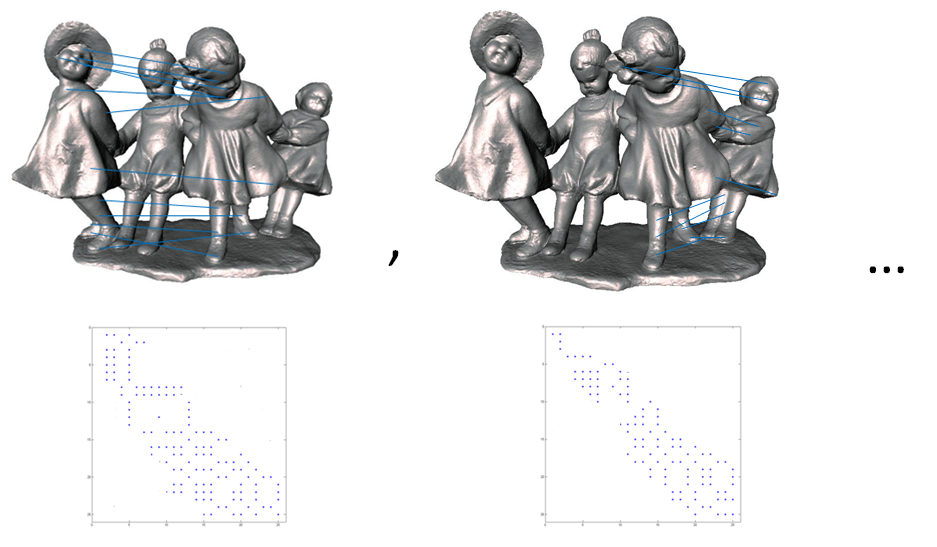}
\end{center}
   \caption{Symmetry extraction in functional space.}
\label{fig:teaser}
\end{figure*}

We present a formal definition of overlapping intrinsic symmetry. An intrinsic symmetry over a shape is a subregion with associated self-homeomorphisms that preserve all pairwise intrinsic distances~\cite{Mitra13}. In this paper, we address the problem of intrinsic symmetry detection and characterization of shapes based on their intrinsic symmetries. Complex shapes often exhibit multiple symmetries that overlap spatially and vary in form as depicted in Figure~\ref{fig:teaser}. Overlapping symmetry analysis enables the construction of high-level representations that enhance the understanding of the underlying shape and facilitate solutions to such problems as shape correspondence, shape editing, and shape synthesis~\cite{Wang11}. However, analysis of overlapping symmetry poses additional challenges as described below. 



Existing approaches to intrinsic symmetry detection, including those based on region growing~\cite{Xu09}, partial matching~\cite{Raviv10}, and symmetry correspondence ~\cite{Lipman10} are not able to extract physically overlapping symmetries.  Lipman et al.~\cite{Lipman10} cluster sample surface points from an input shape and use a symmetry correspondence matrix (SCM) to identify intrinsic symmetry properties of groups of surface points. In their approach, each SCM entry measures how symmetric two surface points are based on some measure of intrinsic geometric similarity between the local neighborhoods of the points. Xu et al.~\cite{xu12} let surface point pairs vote for their partial intrinsic symmetry and perform intrinsic symmetry grouping using a 2-step spectral clustering procedure. However, their approach lacks the ability to retrieve the final symmetry map which makes characterization of the specific intrinsic symmetry a difficult problem.


Our key idea behind our approach to intrinsic symmetry detection and characterization is to approach the problem from a shape correspondence perspective and generate the transformation map which can be further used to describe the symmetry space. To this end, we perform two stages of processing where in the first stage, representative symmetric point pairs are identified based on their local geometry and a global distance representation and in the second stage the original transformation is retrieved as a {\it map} to facilitate further characterization of the underlying symmetry. The detected intrinsic symmetries are quite general, as shown in Figure~\ref{fig:teaser}.

\subsection{Symmetric point pairs}

A fundamental question in symmetry detection is quantifying the extent of symmetry between a pair of points. The primary challenge in identifying potentially symmetric point pairs is to come up with conditions strong enough to adequately constrain the symmetry search space such that the symmetry detection procedure is computationally tractable. We rely on a local criterion, such as geometric similarity, and a global criterion, such as the distance-based symmetry support received by a point pair $\{a, b\}$, to detect and quantify the extent of symmetry between points $a$ and $b$. We refer to a point pair $\{a, b\}$ which satisfies the local geometric similarity criterion as a {\it good voter}. The point pairs within the population of {\it good voters}, that enjoy sufficiently strong global distance-based symmetry support are deemed to be {\it symmetric} point pairs. The global symmetry support for point pair $\{a, b\}$ is quantified by the number of other point pairs which potentially share the intrinsic symmetry properties of  $\{a, b\}$.  The global symmetry support is computed using a simple, distance-based symmetry criterion defined over two point pairs within the population of {\it good voters} as shown in Figure~\ref{fig:overview}.

\subsection{Overview}

\begin{figure*}
\begin{center}
   \includegraphics[width=150mm]{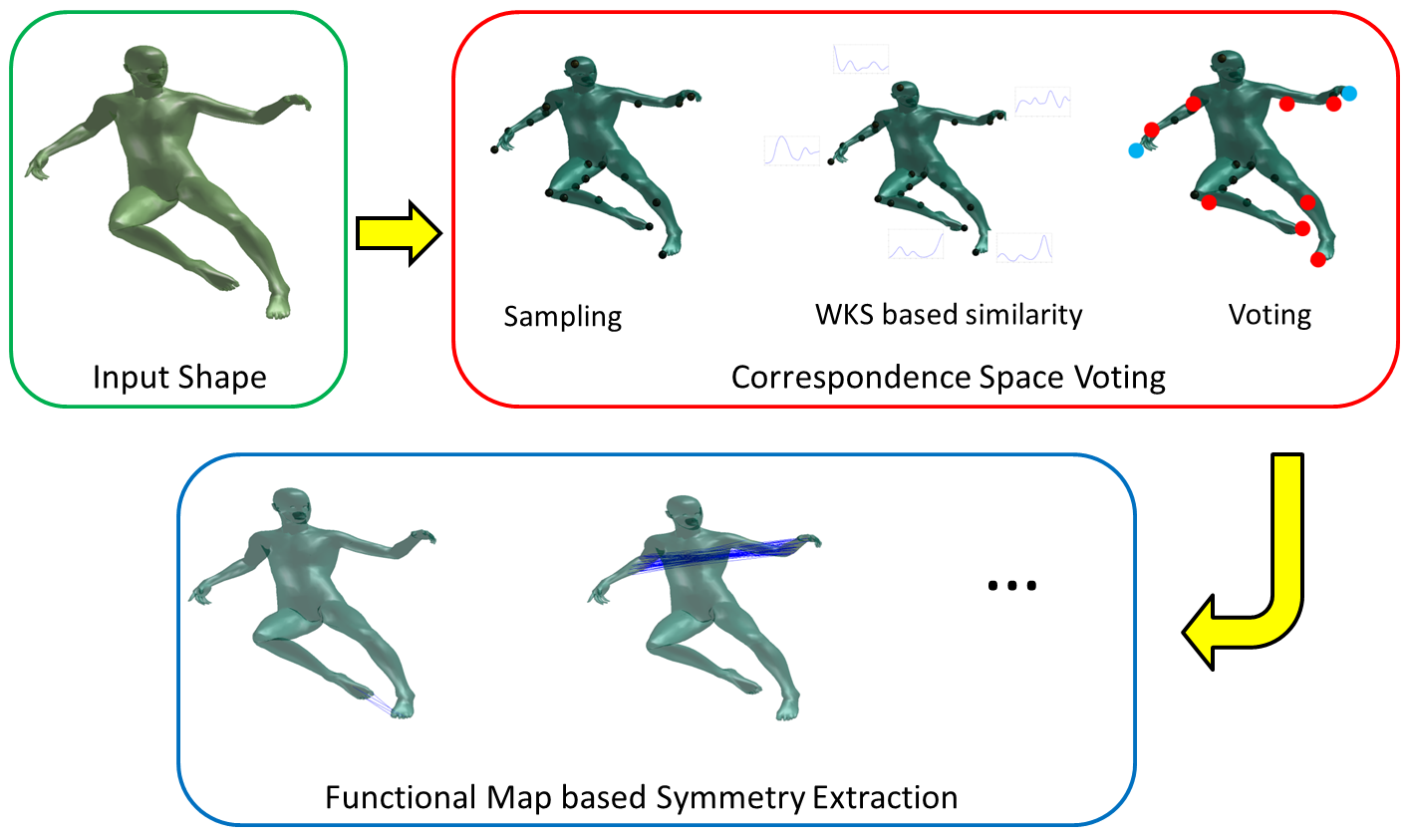}
\end{center}
   \caption{Overview of the proposed symmetry detection and characterization framework.}
\label{fig:overview}
\end{figure*}

The input to the proposed symmetry detection and characterization algorithm is a 3D shape that is approximated by a 2-manifold triangular mesh. The local intrinsic geometry is quantified using the {\it wave kernel signature} (WKS)~\cite{Aubry11} whereas the global intrinsic geometric distance measure chosen is the {\it biharmonic distance measure} (BDM)~\cite{Lipman10a}. The proposed algorithm consists of a voting procedure on the correspondence space defined over a set of locally symmetric surface point pairs sampled from the input 3D shape, followed by the generation of a functional map (see Figure~\ref{fig:overview}). 


\paragraph{Correspondence Space Voting}

The input to the correspondence space voting procedure comprises of point pairs that can be considered as candidates for symmetry. Prior to performing the transformation space mapping procedure, we sample a set of locally symmetric point pairs from the input shape based on the similarity of their WKS's. To estimate the symmetry support received by a point pair, we perform a voting procedure by counting the number of{ \it good voters}
which potentially share the same intrinsic symmetry as the source point pair. The voting procedure ensures that we have a set of good point pair initializations from which we can create an initial map of the symmetry transformation that can then be extrapolated to other surface point pairs on the 3D shape.


\paragraph{Computation of the Functional Map}

A functional map provides an elegant representation for the maps between surfaces, allowing for efficient inference and manipulation. In the functional map approach, the concept of a map is generalized to incorporate correspondences between real valued functions rather than simply between surface points on the 3D shapes. Our choice of the multi-scale eigenbasis of the Laplace-Beltrami operator makes the functional map representation both, very compact and yet informative. We show that our functional map formulation not only results in a compact description of the underlying map of the symmetry transformation, but also enables meaningful characterization of the symmetry transformation.

The primary contributions of our paper can be summarized as follows: 
\begin{enumerate}

\item Robust and meaningful characterization of the symmetry transformation via formulation of a symmetry space to quantitatively distinguish between instances of simple and complex symmetry. 

\item Leveraging of the functional map representation to succesfully represent the map of the underlying symmetry transformation regardless of its complexity. 

\item Providing a solution to the symmetric flipping issue via interpolation of the correspondence from the base surface point pairs using functional maps.

\item Enabling recovery of the symmetry groups via clustering on the functional maps.


\end{enumerate}

\section{Related Work} \label{sec:relatedwork}

The research literature on symmetry detection has grown substantially in recent years as shown in the excellent survey by Mitra et. al.~\cite{Mitra13}. In this paper, we do not attempt to provide an {\it exhaustive} exposition of the state of the art in symmetry detection; rather we focus on discussing existing works that are most closely related to our proposed approach.

\subsection{Symmetry detection in transformation space} 

Several recent approaches to detect approximate and partial extrinsic symmetries have focused on algorithms that cluster votes for symmetries in a parameterized “transformation space” ~\cite{Imiya99};~\cite{Mitra06};~\cite{Yip00}; ~\cite{Li05}]. Mitra et al.~\cite{Mitra06} generate “votes” in a transformation space to align pairs of similar points and then cluster them in a space spanned by the parameters of the potential symmetry transformations. Regardless of how good the shape descriptors are, the aforementioned methods are not effective at finding correspondences between points in complex symmetry orbits that are spread across multiple distinct clusters in the transformation space. Since the dimensionality of the transformation space increases with the complexity of the symmetry, the voting procedure in transformation space becomes increasing intractable when dealing with complex symmetries.

\subsection{Symmetry representation in transformation space} 

There exists a body of published research literature that characterizes shape representations based on the extent of symmetry displayed by an object with respect to multiple transformations. Kazdhan et al. ~\cite{Kazdhan03} have proposed an extension of Zabrodsky’s symmetry distance to characteristic functions, resulting in a {\it symmetry descriptor} that measures the symmetries of an object with respect to all planes and rotations through its center of mass. Podolak et al.~\cite{Podolak06} have extended the symmetry descriptor to define a planar reflective symmetry transform (PRST) that measures reflectional symmetries with respect to all planes through space. Rustamov et al.~\cite{Rustamov08} have extended the PRST to consider surface point-pair correlations at multiple radii. 
Although the above representations provide a measure of symmetry for a regularly sampled set of transformations within a group, they are practical only for transformation groups of low dimensionality (for example, rigid body transformations would require one to store a six-dimensional matrix) and break down when faced with groups of higher dimensionality. 


\subsection{Discovery of repeating structures} 

The exists a class of techniques that exploits the redundancy in repeating structures to robustly detect symmetries~\cite{Bokeloh09};~\cite{Li06};~\cite{Liu07};~\cite{Pauly08};~\cite{Shikhare01}. The transformation space voting method proposed by Mitra et al.~\cite{Mitra06} is extended in~\cite{Pauly08} by fitting parameters of a transformation generator to optimally register the clusters in transformation space.
Berner et al.~\cite{Berner08} and Bokeloh et al.~\cite{Bokeloh09} have taken a similar approach using subgraph matching of feature points and feature lines, respectively, to establish potential correspondences between repeated structures.  This is followed by an iterative closest points (ICP) algorithm to simultaneously grow corresponding regions and refine matches over all detected patterns, allowing the detection of repeated patterns even in noisy data~\cite{Pauly08}, but at the cost of requiring a-priori knowledge of the commutative group expected in the data. Also, the non-linear local optimization procedure within the ICP algorithm could cause it to get trapped in a local minimum if the initialization is not good enough.

\subsection{Eigen-analysis methods} 

Lipman et al.~\cite{Lipman10} have proposed an eigen-analysis technique for symmetry detection that relies on spectral clustering. The top eigenvectors of their geometric similarity-based SCM characterize the symmetry-defining orbits, where each orbit includes all points that are symmetric with one another. However, their work is not suited for multi-scale partial symmetry detection. First, expressing local point similarities as symmetry invariants is only appropriate for global intrinsic symmetry detection. In the case of partial symmetry detection, it is not always possible to reliably judge if two surface points are symmetric by comparing only their point (i.e., local) signatures, especially when one point lies on the boundary of symmetric regions. Moreover, their single-stage clustering procedure is unable to identify overlapping symmetries. Xu et al.~\cite{xu12} have extended the eigen-analysis approach of Lipman et al.~\cite{Lipman10} by incorporating the concept of global intrinsic distance-based symmetry support accompanied by a 2-stage spectral clustering procedure to distinguish between scale detection and symmetry detection. Although they showed some interesting results, the 2-stage spectral clustering procedure made their method extremely slow. Furthermore, the absence of transformation map retrieval meant that  further processing of the detected symmetries, which are represented as point pairs, was extremely inefficient.

The proposed scheme is decoupled into two steps of {\it Correspondence Space Voting} and {\it Transformation Space Mapping}.  The {\it Correspondence Space Voting} technique is inspired from the work of Xu et al.~\cite{xu12}, but in our technique, we bypassed the 2 particularly lengthy steps of spectral clustering and the all-pair geodesic distance calculation to improve the running time quite significantly. Moreover, our introduction of {\it Transformation Space Mapping} in symmetry detection is quite novel in the sense that this not only provides a concise description of the underlying symmetry transformation, but according to our knowledge this is one of the first works which has the unique ability of characterizing the symmetric transformation.

\section{Theoretical Framework} \label{sec:theory}

In this section, we present a formal description of the problem being addressed in this paper. In particular, we define the input to and output of the proposed algorithm. We also define the type of intrinsic symmetries the proposed algorithm is designed to detect and the formal characterization of these symmetries. There are two primary aspects to the theoretical framework for the proposed algorithm, i.e.,  {\it correspondence space voting} (CSV) and {\it functional map retrieval} (FMR). In the case of CSV, a joint criterion, that combines local intrinsic surface geometry and global intrinsic distance-based symmetry, is proposed and shown to result in a provably necessary condition for intrinsic symmetry. 
In the case of FMR,  a formal scheme for the characterization of the detected symmetries, based on their complexity is proposed. 

We have mainly restricted our study of intrinsic symmetries to isometric involutions for two reasons; first, having a provably necessary condition for intrinsic symmetry provides theoretical soundness and second, the proposed symmetry criterion bounds the search space sufficiently, ensuring that the solution is computationally tractable~\cite{xu12}.
Following the initial detection of good symmetric correspondences, we proceed to retrieve the map of the symmetry transformation by leveraging the functional map framework. Moreover, the intrinsic symmetry ensures that the retrieved maps exhibit a diagonality characteristic. A cost matrix is designed to exploit this characteristic such that the inner product of functional map with the cost matrix results in a {\it quantitative} evaluation of the complexity of the detected symmetry.

\subsection{Input to the Proposed Algorithm}

Our problem domain is a compact, connected 2-manifold, $M$, with or without a boundary. The manifold $M$ is a {\it 3D shape},  i.e., $M \rightarrow \mathbb{R}^3$. Distances on the manifold $M$ are expressed in terms of an intrinsic distance measure. In particular, the biharmonic distance measure (BDM) is used on account of its ease of calculation and greater robustness to local surface perturbation when compared to the geodesic distance measure $d_M$~\cite{Lipman10a}. 
In the remainder of the paper, we use the term intrinsic distance and biharmonic distance interchangeably.


\subsection{Output of the Proposed Algorithm}

In the proposed algorithm each symmetry transformation is represented by a map. Consequently, the output of the proposed algorithm consists of maps represented as matrices.  The output can be regarded as a complete description of all the overlapping intrinsic symmetries represented in a compact and informative manner.

\subsection{Defintion of Intrinsic Symmetry}

Suppose we are given a compact manifold $M$ without boundary. Following~\cite{Raviv07} we call $M$ intrinsically symmetric if there exists a homeomorphism $T \colon M \rightarrow M$ on the manifold that preserves all geodesic distances. That is:


\begin{equation} 
d_M(p,q) = d_M(T(p),T(q))  \forall p,q \in M \label{eq:Symmetry}
\end{equation}
where $d_M(p,q)$ is the intrinsic distance between two points on the manifold. In this case, we call the mapping $T$ an intrinsic symmetry.


\subsection{Symmetry Criteria}

We propose two simple criteria to test whether two surface point pairs on the manifold potentially share the same intrinsic symmetry. Specifically, given two surface point pairs $\{x, x^{\prime}\}$ and $\{y, y^{\prime}\}$ on manifold $M$, the first criterion, which is based on local intrinsic geometry, determines the symmetry potential of the two surface point pairs by comparing their corresponding wave kernel signatures (WKS's) as follows:
\begin{eqnarray}
WKS(x,t) & = & WKS(x^{\prime},t) \; {\rm and} \nonumber \\
WKS(y,t) &  = & WKS(y^{\prime}, t) \; \forall t \geq  0 \label{eq:WKScriterion} 
\end{eqnarray}
where $t$ is a scale parameter. 
The second criterion is based on intrinsic distance as follows: 
\begin{eqnarray} 
d_M(x,y) & = & d_M(x^{\prime},y^{\prime}) \; {\rm and} \nonumber  \\
d_M(x,y^{\prime}) & = & d_M(y,x^{\prime}) \label{eq:DistCriterion}
\end{eqnarray}
The above two criteria are necessarily satisfied if the surface point pairs under consideration correspond to the same intrinsic symmetry.

\subsection{Limitations}

\begin{figure}
\begin{center}
   \includegraphics[width=85mm]{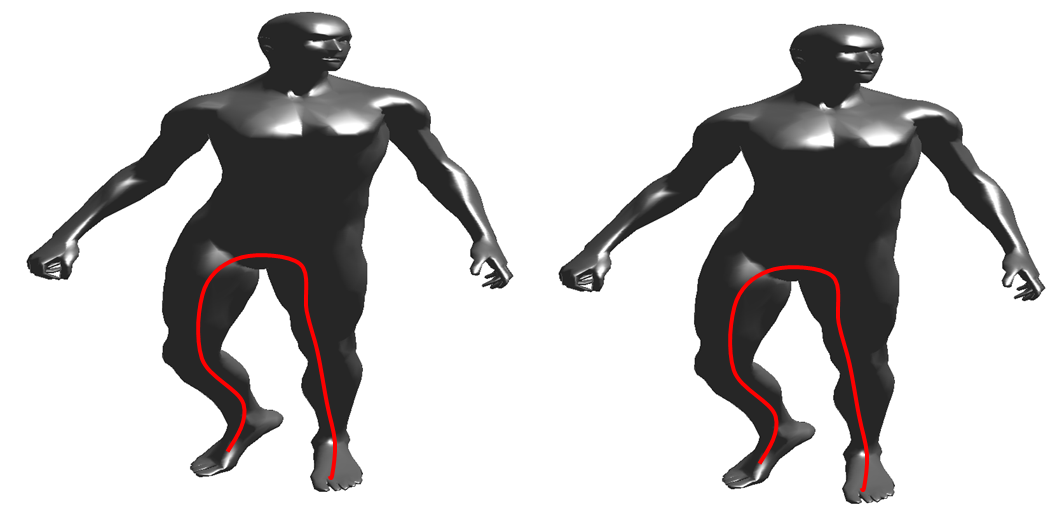}
\end{center}
   \caption{Limitation of the proposed symmetry detection algorithm.}
\label{fig:failure}
\end{figure}

The proposed framework fails to detect intrinsic symmetry in cases where the second symmetry criterion (in equation \eqref{eq:DistCriterion}) is not satified. For example, Figure \ref{fig:failure}, depicts two human figures that form what may be perceived as a translational symmetry, although they do not possess intrinsic symmetry. The second symmetry criterion (equation \eqref{eq:DistCriterion}) fails to hold in this situation. Generally speaking, the proposed  algorithm is not designed to detect all forms symmetry resulting from repeated patterns, especially if the patterns are not connected. In contrast to the approach of Xu et al.~\cite{xu12}, the use of biharmonic distance instead of geodesic distance as the intrinsic distance measure ensures that the proposed algorithm is capable of detecting intrinsic symmetry even in the presence of small perturbations (such as small bumpy regions) on the 3D surface~\cite{Lipman10a}. 

\section{Correspondence Space Voting}

The first step in the proposed symmetry detection and characterization framework is the correspondence space voting (CSV) procedure. Although a voting procedure has been previously incorporated in an earlier symmetry detection technique, it was carried out primarily in transformation space~\cite{Mitra06}. The importance of correspondence space for the detection of symmetry was explained more recently by Lipman et al.~\cite{Lipman10}.  In our case, although the detected symmetry is finally represented in functional space, it is critical to have good initialization to ensure the success of the final map generation. We have designed and implemented a CSV algorithm to facilitate good initial guesses. The CSV algorithm comprises of three stages described in the following subsections:


\subsection{Point Selection}

\begin{figure}
\begin{center}
   \includegraphics[width=85mm]{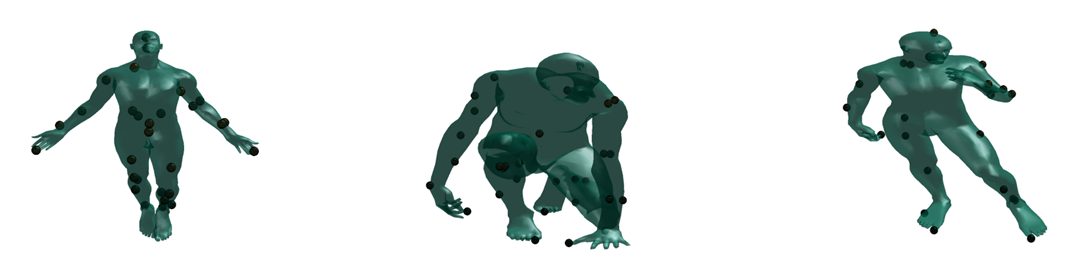}
\end{center}
   \caption{Sampling results on three different {\it human} shapes}
\label{fig:sampling}
\end{figure}

A subset of points with adequate discriminative power needs to be selected prior to the generation of surface point pairs. A subset $X$ consisting of $n$ sample points is chosen from the surface of the given input 3D shape using the {\it Farthest Point Sampling} strategy~\cite{Eldar97}.  Although originally designed to operate on geodesic distances generated using the marching cubes algorithm, in our case we have employed the farthest point random sampling strategy in biharmonic distance space. The results of the sampling procedure are depicted in Figure~\ref{fig:sampling}. 

The {\it biharmonic distance measure} (BDM) is similar in form to the {\it diffusion distance measure} (DDM)~\cite{Ling06}. 
The BDM kernel is based on the Green's function of the biharmonic differential equation. In the continuous case, the (squared) biharmonic distance between two points $x$ and $y$ can be defined using the eigenvectors ($\phi_k$) and eigenvalues ($\lambda_k$) of the Laplace- Beltrami operator~\cite{Lipman10} as follows:

\begin{equation}
d_M(x, y)^{2} = \sum_{k=1}^\infty \dfrac{(\phi_k(x) - \phi_k(y))^2} {\lambda_k^2} \label{eq:BDM} 
\end{equation}

The above definition of the BDM is slightly different from that of the DDM where the denominator in the case of the DDM is \begin{math} e^{2t\lambda_k} \end{math}. However, this subtle change ensures greater control over the characterization of the global and local properties of the underlying manifold in the case of the BDM. Consequently, the BDM is fundamentally different from the DDM with significantly different properties. 

The BDM, as expressed in equation \eqref{eq:BDM}, captures the rate of decay of the normalized eigenvalues  \begin{math} \lambda_k \end{math} of the Laplace- Beltrami operator;  if the decay is too slow, it produces a logarithmic singularity along the diagonal of the Green's function~\cite{Yen07}. Alternatively, too fast a decay basically ignores eigenvectors associated with higher frequencies, resulting in the BDM being global in nature (i.e., the local surface details are ignored). Lipman {\it et al.}~\cite{Lipman10a}, demonstrated that performing  quadratic normalization provides a good balance, ensuring that the decay is slow enough to capture the local surface properties around the source point and yet rapid enough to encapsulate global shape information.

In particular, Lipman {\it et al.}~\cite{Lipman10a} have theoretically proved two important properties of the BDM, i.e., that it is  (i)  a metric, and (ii) smooth everywhere except at the source point where it is continuous. The key observation is that for 3D surfaces, the eigenvalues $\lambda_k$, $k = 1, 2, \ldots$, of the Laplacian are an increasing function of $k$ resulting in the continuity of the BDM everywhere and also smoothness of the BDM everywhere except at the source point, where it has only a derivative discontinuity.

In our implementation of the farthest point random sampling strategy, a single point is selected randomly at first and the remaining points are chosen iteratively from remainder set by selecting the farthest point in the biharmonic distance space at each iteration. This strategy generates a set of points located mostly in the vicinity of the shape extrema which can then be used in the subsequent surface point pair generation procedure.

\subsection{Surface Point Pair Generation}

From the chosen subset $X$ consisting of $n$ sample points, the surface point pairs are generated by exploiting the similarity of their local intrinsic geometric structure. The similarity of local intrinsic geometric structure of two surface points is determined by comparing their corresponding wave kernel signatures (WKS's).  To determine the WKS of a surface point $x$, one evaluates the probability of a quantum particle with a certain energy distribution to be located at point $x$. The behavior of the quantum particle on the surface is governed by the Schrodinger equation. 
Assuming that the quantum particle has an initial energy distributed around some nominal energy with a probability density function $f(e)$, the solution of the Schrodinger equation can then be expressed in the spectral domain as:
\begin{equation}
\psi(x,t)=\sum_{k>=1} e^{ie_kt}f(e_k) \phi_k(x) \label{eq:Schrodinger} 
\end{equation}

Aubry et al.~\cite{Aubry11} considered a family of log-normal energy distributions centered around some mean log energy $\log e$ with variance $\sigma^2$. This particular choice of distributions is motivated by a perturbation analysis of the Laplacian spectrum~\cite{Aubry11}. 
Having fixed the family of energy distributions, each point $x$ on the surface is associated with a WKS of the form: $p(x) = (p_{e_1} (x), \ldots, p_{e_n}(x))^T $ 
where $p_{e_i}(x)$ is the probability of measuring a quantum particle with the initial energy distribution $e_i (x)$ at point $x$. Aubry et al. use logarithmic sampling to generate the values $e_1 (x), \ldots, e_n (x)$.

\begin{figure*}
\begin{center}
   \includegraphics[width=150mm]{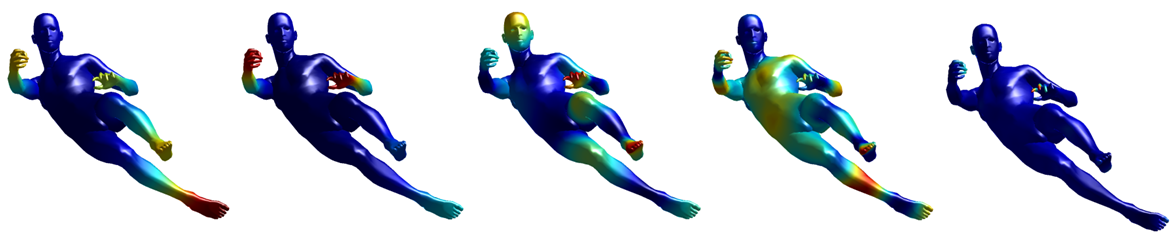}
\end{center}
   \caption{The band-pass characteristic of the WKS at different scales for the same {\it human} figure.}
\label{fig:wks}
\end{figure*}

The WKS can be shown to exhibit a band-pass characteristic. This reduces the influence of low frequencies and allows better separation of frequency bands across the descriptor dimensions. As the result, the wave kernel descriptor exhibits superior feature localization compared to the Heat Kernel Signature (HKS)~\cite{Litman13}.
Although the WKS is invariant under isometric deformation, in most practical cases where the underlying surface is represented by a discrete triangular mesh, it is not possible to strictly satisfy the invariance criterion in equation \eqref{eq:WKScriterion}.  Consequently, we have considered the 2-norm (simply squared Eucledian distance in the WKS space) of the WKS and chosen surface point pairs from the subset $X$ which satisfy the WKS invariance criterion in equation \eqref{eq:WKScriterion} to within a prespecified  threshold $T_{WKS}$ instead of requiring strict equality. The surface point pairs that satisfy the WKS invariance criterion in equation \eqref{eq:WKScriterion} are considered for the next step of global distance-based voting. This relaxation of the WKS invariance criterion ensures that even the overlapping intrinsic symmetries are considered for voting. We have coined the term {\it good voters} to denote the subset of surface point pairs that satisfy the WKS invariance criterion (to within the prespecified threshold). 

\subsection{Global Distance-based Voting}

The global distance-based voting step in the proposed symmetry detection technique is inspired by the work of  Xu et. al.~\cite{xu12}. A subset of symmetric point pairs is extracted from the set of {\it good voters} using the global distance-based voting procedure prior to functional map generation. The goal of the voting procedure is to accumulate symmetry support for the {\it good voters} and extract the symmetric point pairs based on the level of symmetry support received. 

A point pair in the set of {\it good voters} is deemed to be symmetric if it has a sufficiently large global symmetry support, which in~\cite{xu12} is measured by the number of point pairs that satisfy the intrinsic distance criterion in equation \eqref{eq:DistCriterion}. Xu et al.~\cite{xu12} have presented a voting technique based on two point pairs followed by spectral clustering on the symmetric point pairs which enables one to distinguish whether a point pair supports one particular symmetry or more than one type of symmetry. However, since we are are not interested, at this stage, in making the above distinction between the point pairs, we have chosen to adopt a straightforward approach of choosing a point pair and letting the set of {\it good voters} vote and decide whether or not the particular point pair satisfies the intrinsic distance criterion in equation \eqref{eq:DistCriterion}. Another modification to the previous voting procedure of Xu et al.~\cite{xu12} is our use of the biharmonic distance as the intrinsic distance instead of the geodesic distance. This modification provides two advantages over the previous voting procedure, First, since the computation of geodesic distances between all surface point pairs is more intensive than the computation of biharmonic distances, our procedure is much faster. Second, biharmonic distances are less sensitive to noise and surface perturbations compared to geodesic distances, making our procedure more robust.

\section{Symmetry Extraction using Functional Maps}

Before describing our functional map approach to symmetry extraction, we first provide an overview of the functional map framework proposed by Ovsjanikov et al.~\cite{Ovsjanikov12}. A functional map is a novel approach for inference and manipulation of maps between shapes that tries to resolve the issues of correspondences in a fundamentally different manner. Rather than plotting the corresponding points on the shapes, the mappings between functions defined on the shapes are considered. This notion of correspondence generalizes the standard point-to-point map since every pointwise correspondence induces a mapping between function spaces, while the opposite, in general, is not true.

The new framework described above provides an elegant way to avoid direct representation of correspondences as mappings between shapes using a functional representation. Ovsjanikov et al.~\cite{Ovsjanikov12} have noted that when two shapes $X$ and $Y$ are related by a bijective correspondence $t: X \rightarrow Y$ , then for any real function $f :X \rightarrow \mathbb{R}$, one can construct a corresponding function $g :Y \rightarrow \mathbb{R}$ as $g :f \circ t^{-1}$. In other words, the correspondence $t$ uniquely defines a mapping between the two function spaces $F(X,\mathbb{R}) \rightarrow F(Y,\mathbb{R})$ , where $F(X,\mathbb{R})$ denotes the space of real functions on $X$. Equipping $X$ and $Y$ with harmonic bases, $\{\phi_i\}_{i \geq 1}$ and $\{\psi_j\}_{j \geq 1}$, respectively, one can represent a function $f : X \rightarrow \mathbb{R}$ using the set of (generalized) Fourier coefficients $\{a_i\}_{i \geq 1}$ as $f=\sum_{i \geq 1}a_i \phi_i$. Translating this representation into the other harmonic basis $\{\psi_j\}_{j \geq 1}$, one obtains a simple representation of the correspondence between the shapes given by $T(f)=\sum_{i,j \geq 1} a_i c_{ij} \psi_j$ where $c_{ij}$ are Fourier coefficients of the basis functions of $X$ expressed in the basis of $Y$, defined as $T(\phi_i )=\sum_{i,j \geq 1} c_{ij}\psi_j$. The correspondence $t$ between the shapes can thus be approximated using $k$ basis functions and encoded using a $k \times k$ matrix $C=(c_{ij})$ of these Fourier coefficients, referred to as the functional matrix. In this representation, the computation of the shape correspondence $t:X \rightarrow Y$ is translated into a simpler task of determining the functional matrix $C$ from a set of correspondence constraints. The matrix $C$ has a diagonal structure if the harmonic bases $\{\phi_i\}_{i \geq 1}$ and $\{\psi_j\}_{j \geq 1 }$ are compatible, which is a crucial property for the efficient computation of the correspondence. 

In our symmetry extraction algorithm, instead of comparing two different shapes, we propose to compare two symmetric regions within the same shape. In particular, based on the previously detected set of symmetric point pairs, we leverage the functional map representation in the following manner: for each pair of symmetric points, we deem one point as the source and the other as the destination and choose a local region around each point. The ordering of the source and destination points within the pair is the same as originally chosen during the voting procedure. The corresponding eigenbases for the points in the source and destination regions are computed. These eigenbases are ordered based on their similarity with each other and the final functional map for that particular symmetry is extracted. The functional map representation ensures that (a) the problem of symmetry extraction is tractable and, (b) the resulting symmetry can be represented, not by a large matrix of point correspondences, but rather as a more compact map which can be further manipulated for other applications as well.  


\section{Symmetry Characterization using Functional Maps}

In general, the characterization of a specific transformation based on its functional map is a challenging task.  
However in our case, the proposed CSV framework ensures that the point pairs used in the generation of the functional map are intrinsically symmetric to a reasonable extent. This property of intrinsic symmetry ensures that the resulting functional map is diagonal or close to diagonal~\cite{Ovsjanikov12}. However, in reality, there are several cases where the actual transformation deviates substantially from $\epsilon$-isometric deformation resulting in a densely populated functional matrix $C$. The diagonality property of the $C$ matrix was first exploited succesfully in an intrinsic correspondence framework using a sparse modeling technique~\cite{Porkass13}. In their paper, the authors introduced a weight matrix $W$ of same size as $C$ where lower weights are assigned to elements close to the diagonal, and larger weights are assigned to the farther off-diagonal elements by using an inverted Gaussian model where the 0 weights are assigned at diagonal and the weights at off-diagonals are larger. The element-wise multiplication of $W$ with $C$ is a determining factor in the assessment of the diagonality of $C$ since the matrix inner product results in higher values for the off-diagonal elements and lower values for the diagonal elements thereby resulting in a measure of the diagonality of the matrix $C$.


\begin{figure}
\begin{center}
   \includegraphics[width=80mm]{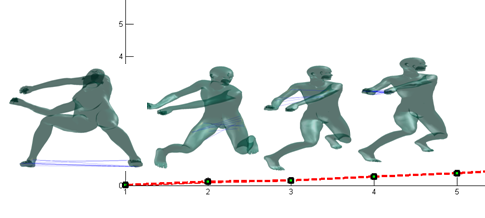}
\end{center}
   \caption{Symmetry characterization based on functional maps, in particular, the complexity of the symmetry transformation is characterized by the wieght matrix $W$ and represented in the increasing order of the value of inner product of $W$ and  $C$.}
\label{fig:charac}
\end{figure}

In the context of symmetry characterization, problem, we assume that the off-diagonality of $C$ corresponds to the complexity of the symmetry transformation and therefore is penalized during multiplication with elements of the weight matrix $W$, i.e., more non-zero off-diagonal elements in $C$ would represent a more complex symmetry transformation. Thus, we can, not only determine the complexity of the symmetry transformation but can also succesfully formulate a 1D semi-metric symmetry space, wherein each symmetry transformation is represented as a point in the symmetry space with a value given by the inner product beween the matrix $C$ and weight matrix $W$. The Euclidian distance between the points in the 1D symmetry space represents the complexity distance between the transformations. In the symmetry space, any perfectly isometric transformation will result a point at origin of the line. On the other hand, more complex symmteric transformations would result in points farther away signifying more complex transformations. This space is called a semi-metric space because of it follows two main properties of the distance definition but not the third one. For two points $X$ and $Y$ in the symmetry space, the distance $d(X,Y)$ from $X$ to $Y$ always follows

\begin{enumerate}

\item Non-negativity: $d(X,Y) \ge 0$

\item Symmetry: $d(X,Y) = d(Y,X)$

but not

\item Triangle Inequality.

\end{enumerate}

It is also possible to cluster the ponts in the 1D symmetry space to identify intrinsic symmetries which are potentially similar in nature as shown in the Experimental Results section to follow.


\section{Experimental Results}

In this section, we present and discuss the results obtained by the proposed intrinsic symmetry detection algorithm on 3D shapes. We also provide comparisons of our results with those obtained from the most closely related approaches ~\cite{xu12},~\cite{Lipman10}. We also shown some applications where the detected symmetries can be further analyzed for symmetry characterization and clustering, potentially revealing greater semantic information about the underlying 3D shape. Most of the 3D shape models used in our experiments are from the {\it Non-rigid World} dataset~\cite{Bronstein07} unless mentioned otherwise.  


\subsection{Preprocessing}

Several discrete schemes have been proposed in recent years to approximate the Laplace-Beltrami operator on triangular meshes. Among these, the one mostly used most widely for computing the discrete Laplace operator is the cotangent (COT) scheme, originally proposed by Pinkall and Polthier~\cite{Pinkall93}. Recently, Belkin et al~\cite{Belkin08} proposed a discrete scheme based on the heat equation, which has been proven to possess the point-wise convergence property for arbitrary meshes. Although it is well known that no discrete Laplace operator can share all of the properties of its continuous counterpart, in our experiments the aforementioned discrete schemes produce eigenfunctions that approximately preserve the convergence property of the continuous Laplace operator for a reasonably well sampled triangular mesh. 
In all of the experiments described below, we have used the cotangent scheme~\cite{Pinkall93} for the computation of the discrete Laplace-Beltrami operator. 

\subsection{Symmetry Detection}

The results of the proposed symmetry detection algorithm are depicted in Figures \ref{fig:teaser}, \ref{fig:overlap}, \ref{fig:noise} and \ref{fig:clustering}. Several important properties of the proposed algorithm are highlighted in these results.

\subsubsection{General Symmetry Detection}

The ability of the proposed algorithm to identify multiple intrinsic symmetries is evident from the results shown in Figures \ref{fig:overlap} and \ref{fig:clustering}. The extracted symmetries are seen to cover the global symmetry of the underlying 3D shape which has undergone approximate isometric deformations. 
Additionally, the proposed algorithm is also observed to be capable of detecting symmetry transformations that cover individual components of a 3D object that possess various forms of self-symmetry.

\subsubsection{Overlapping Symmetry Detection}

\begin{figure}
\begin{center}
   \includegraphics[width=80mm]{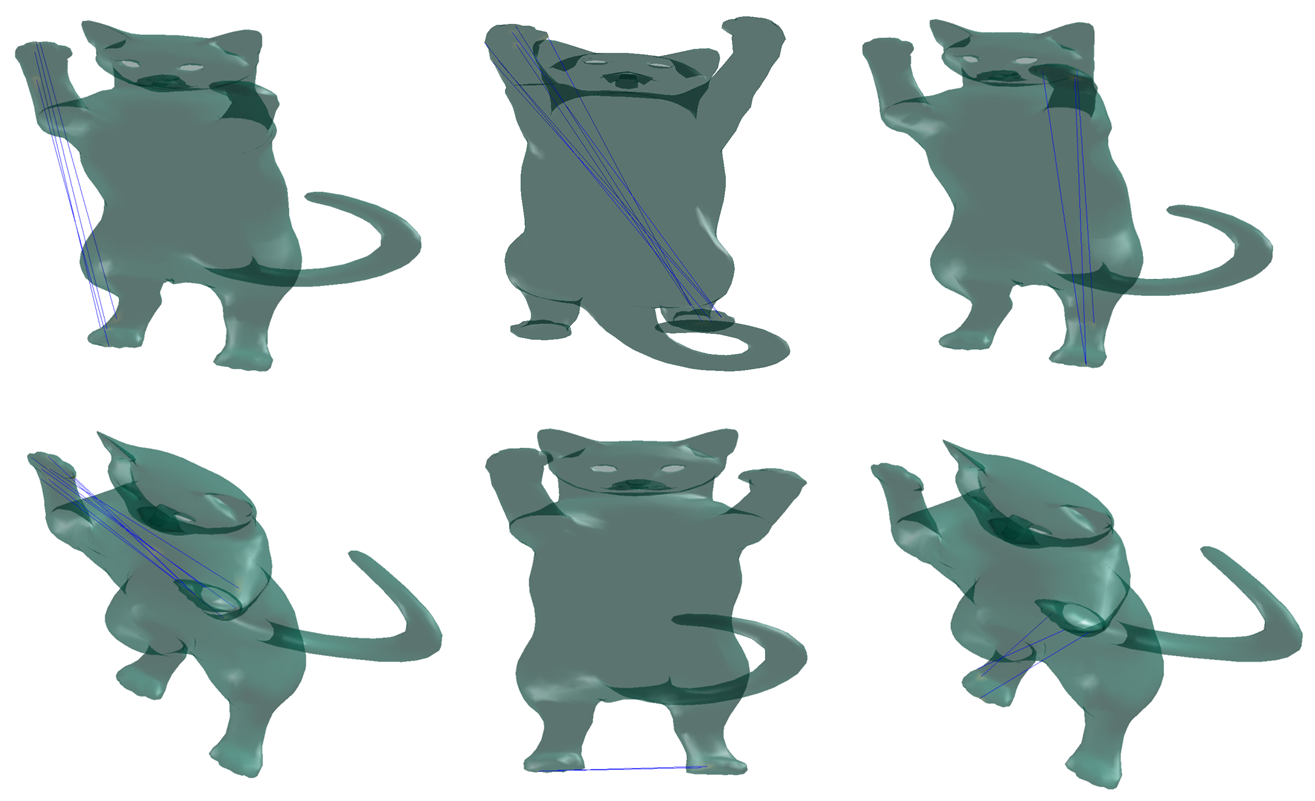}
\end{center}
   \caption{Overlapping symmetry detection on the {\it Cat } shape model.}
\label{fig:overlap}
\end{figure}

One particularly important aspect of the proposed algorithm is its ability to detect instances of overlapping symmetry. An instance of overlapping symmetry is deemed to occur when a specific region on the surface of the 3D shape is simultaneously subjected to more than one symmetry transformation and, as a result,  is symmetric to more than one region on the 3D shape surface. For example, Figure \ref{fig:overlap} shows that the overlapping symmetry between {\it all} paws of the {\it Cat} shape model is succesfully detected by the proposed technique. Six different combinations of symmetry transformation between the four paws are depicted in Figure \ref{fig:overlap}.

\subsubsection{Noise} 
\begin{figure}
\begin{center}
   \includegraphics[width=85mm]{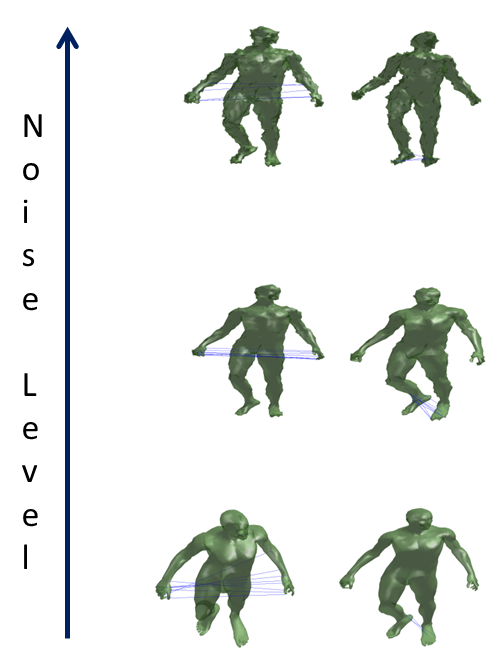}
\end{center}
   \caption{Robustness of symmetry detection for three different levels of synthetic white Gaussian noise added to the {\it Human} shape model.}
\label{fig:noise}
\end{figure}

Although the CSV procedure is statistical in nature, the proposed biharmonic distance-based voting scheme ensures its robustness to noise. In particular, Figure \ref{fig:noise} demonstrates the robustness of the proposed symmetry detection technique for different levels of synthetic Gaussian noise added to the {\it Human} shape model. 

\subsection{Performance Statistics}

All of the experiments reported in this paper were performed on an Intel Core$^{\texttrademark}$ 3.4 GHz machine with 8 GB RAM. For all the example models, the number of sample points were in the range [20, 100]. Table \ref{tab:timing} reports the timing results for the various steps in the proposed symmetry detection algorithm. In particular, unlike~\cite{xu12}, wherein the most time consuming step of all-pairs geodesic distance computation is not reported, we also report the timing results for the equivalent step in our formulation i.e., the all-pairs biharmonic distance computation. The most time consuming step in the proposed algorithm, i.e., the all-pairs biharmonic distance computation, accounts for around 80\% of the execution time of the proposed algorithm. More importantly, bypassing the two-step spectral clustering procedure described in~\cite{xu12} reduces significantly the computation time of the proposed algorithm.

\begin{table}
\tbl{Timing results for the various steps in the proposed algorithm.}{
    \begin{tabular}{| c | c | c | c | c |}
    \hline
    Points & Biharmonic & FMap & Extraction & Total \\ \hline
    3400 & 32 & 2 & 5 & 39\\ \hline
    10000 & 41 & 3 & 7 & 51\\ \hline
    50000 & 237 & 4 & 15 & 256\\
    \hline
    \end{tabular}}
 \label{tab:timing} 
\end{table}


\subsection{Comparisons}

We have compared the proposed symmetry detection algorithm primarily with methods that could be deemed sufficiently similar~\cite{xu12},~\cite{Lipman10}. The symmetry detection technique of Xu et al.~\cite{xu12}  can detect overlapping partial intrinsic symmetries whereas that of  Lipman et al.~\cite{Lipman10} is designed to deal with partial extrinsic symmetries. However, if the symmetric sub-shapes do not undergo significant pose variations, the global alignment component of ~\cite{Lipman10} may allow it to detect certain partial intrinsic symmetries as well. However, whereas both methods~\cite{xu12},~\cite{Lipman10} are capable of detecting instances of partial intrinsic symmetry, neither is able to characterize the underlying symmetry. 

\begin{figure}
\begin{center}
   \includegraphics[width=80mm]{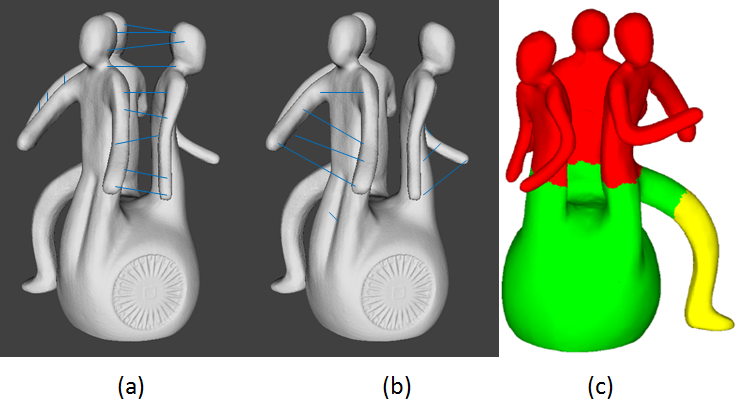}
\end{center}
   \caption{In comparison of (c) the symmetry-factored embedding (SFE) technique of Lipman et al.~\protect\cite{Lipman10}, which primarily detects instances of global intrinsic non-overlapping symmetry, the proposed CSV procedure ensures the detection of instances of overlapping intrinsic symmetry (a,b) as well.}
\label{fig:sfe}
\end{figure}

\begin{figure}
\begin{center}
   \includegraphics[width=80mm]{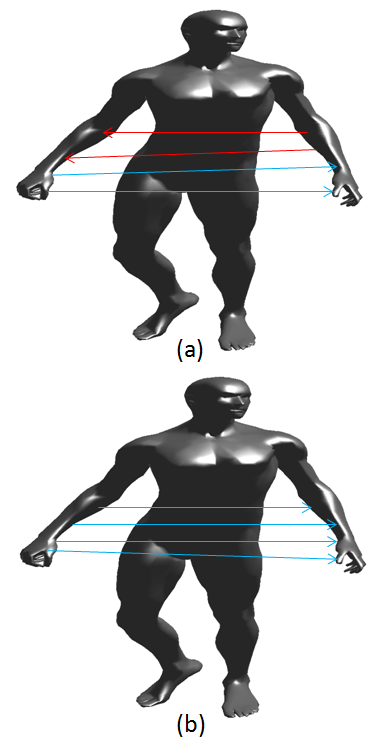}
\end{center}
 \caption{In contrast to the technique of Xu et al.~\protect\cite{xu12} (a), which relies solely on global intrinsic distance and consequently, fails to detect symmetric flips (in red), the proposed functional map-based interpolation of the symmetric correspondence (b) ensures robust detection of symmetric flips.}
\label{fig:flip}
\end{figure}

In contrast, the proposed algorithm, not only detects overlapping intrinsic symmetries, but it also has the ability to characterize and cluster the detected symmetries in symmetry space. Since their entire formulation is based solely on global intrinsic distance-based voting, the technique of Xu et al.~\cite{xu12} suffers from the shortcoming of not being able to reliably detect symmetry flips. The symmetry flip phenomenon is deemed to occur when for the same symmetric transform, the point correspondences interchange their relative positions due to ill-posed symmetry detection criteria. The proposed algorithm, on the other hand, interpolates the functional map of symmetry transformations from the chosen directed point pair, to the remaining point pairs, i.e., once the source and destination within a point pair are  identified, the remaining correspondences are obtained via  interpolation using the functional map. The functional map ensures that the interpolated point pairs observe the same direction of symmetry as the initial directed pair, i.e., the functional map preserves the relative positions of points within a point pair that are subject to the same symmetry transformation. The functional map thus provides a natural solution to the symmetry flip problem as evidenced in Figure~\ref{fig:flip}. The symmetry-factored embedding (SFE) technique of Lipman et al.~\cite{Lipman10}, though designed for extrinsic symmetry detection, in simpler cases, is able to detect intrinsic symmetries as well. However, it fails completely in cases of overlapping symmetries whereas the proposed CSV procedure ensures the detection of instances of overlapping intrinsic symmetry as shown in Figure~\ref{fig:sfe}.


\subsection{Quantitative Evaluation}

\begin{figure}
\begin{center}
   \includegraphics[width=80mm]{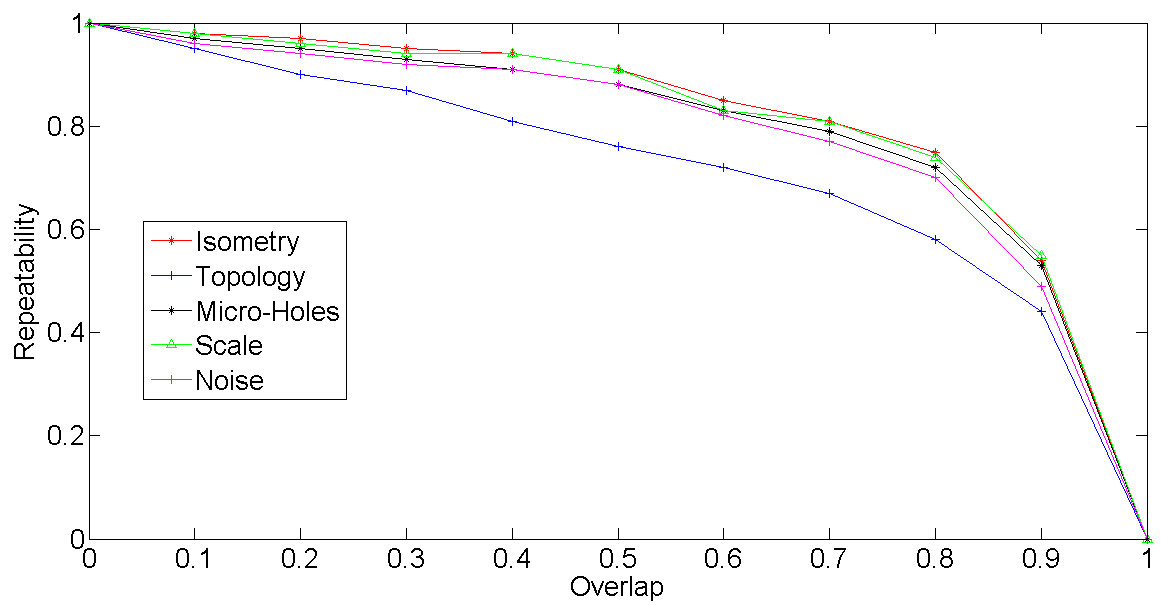}
\end{center}
   \caption{Overlap vs. Repeatability plot for the proposed symmetry detection technique on the SHREC 2010 benchmark dataset.}
\label{fig:quant}
\end{figure}

In this section, we present quantitative evaluation of the results obtained by the proposed approach. In order to evaluate the proposed algorithm, we used the SHREC 2010 feature detection and description benchmark~\cite{Bronstein10}. This dataset comprises of three shapes (null shapes) and a set of shapes obtained by applying a set of transformations on the null shapes. Shapes have approximately 10,000 to 50,000 vertices and their surfaces are represented by triangular meshes. We have specifically considered shapes which have undergone changes characterized by isometry, topology, micro-holes, scale and noise. For each transformation, in the initial phase, a total 50 sample points are generated using the farthest random sampling strategy as shown in Figure~\ref{fig:quant}. Almost all transformations maintain a high repeatability (greater than 80\%) at overlap values $\le 0.75$ except for changes in topology.

\subsection{Symmetry Characterization}

The diagonality property of the functional map could be used to characterize the underlying symmetry transformation and  classify it as a simple transformation or as one that is more complex in nature. Our assumption is that, greater the complexity of symmetry transformation, greater the deviation of the shape deformation from intrinsic isometry, resulting in a deformation characterized by a functional matrix $C$  with higher off-diagonal element values. The resulting  characteriztion of the isometric deformation is depicted in Figure~\ref{fig:charac}.

\subsection{Symmetry Group Retrieval}

\begin{figure*}
\begin{center}
   \includegraphics[width=150mm]{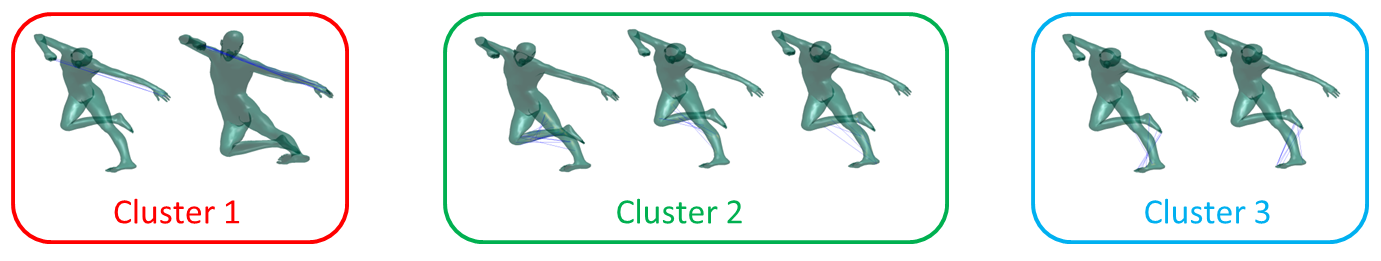}
\end{center}
   \caption{Clustering of the functional maps resulting in three symmetry groups.}
\label{fig:clustering}
\end{figure*}

As discussed in~\cite{xu12}, symmetry has a group structure. Consequently, the retrieval of symmetry transformation can be cast as a clustering problem from an algorithmic perspective. We exploit the functional maps generated previously to cluster the detected instances of symmetry in transformation space and retrieve the symmetry groups. Due to the symmetric flip problem inherent in the CSV procedure, we adjusted the relative positions of source and destination points within each point pair by finding the minimum distance between the source and destination points in the biharmonic distance space. This ensures that the functional maps generated from potentially similar transformations will have similar structure since the symmetry flip problem between the maps is resolved. In particular, we have used a simple $k$-means clustering algorithm to cluster the functional maps based on their symmetry groups as depicted in Figure~\ref{fig:clustering}.

\section{Conclusion and Future Directions}

We have presented an algorithm for detection and characterization of intrinsic symmetry in 3D shapes. While the results obtained are encouraging, we regard our work as an initial attempt towards complete understanding of and a verifiable solution to the general problem of symmetry detection and characterization. We have identified some limitations of our approach and we hope to address these in our future work.

To the best of our knowledge, this is one of the first attempts to formalize the symmetry analysis problem not only as one of symmetry detection, but as one that can be extended to include symmetry characterization and symmetry clustering in the transformation space. In particular, the introduction of the functional map formalism in symmetry detection enables us to come up with a novel representation of the symmetry transformation as a map. In future, we aim to formulate operations, such as addition and subtraction, on these generated maps that would potentially provide a deeper and more comprehensive understanding of intrinsic symmetry in general. 

The incorporation of {\it Transformation Space Mapping} in Symmetry characterization is a completely new idea and the full potential of it can only be realized after more extensive experimentations. In particular, we plan to study the possibility of map based exploration of similar symmetric transformations across shapes in near future. Another important direction that can be considered is the possibility of incorporating this technique in Computer Aided Geometric Design for urban architecture. In urban architecture, symmetric repetition of same pattern is a common thing and during the design phase, if the basic structure is stored only once and the symmetric repetitions are saved as {\it Functional Maps}, it can possibly solve both the space complexity problem and the design efficiency problem. 


\end{document}